\documentclass[preprint]{aastex61}

\usepackage{rotating}

\usepackage{amsmath,amsbsy}
\usepackage{graphicx}

\shorttitle{Murrili Fall}
\shortauthors{Sansom et al.}

\begin{document}
\title{Murrili meteorite's fall and recovery from Kati Thanda}

\correspondingauthor{Eleanor K. Sansom}
\email{eleanor.sansom@curtin.edu.au}

\author[0000-0003-2702-673X]{Eleanor K. Sansom}
\affiliation{School of Earth and Planetary Sciences, Curtin University, Bentley, WA 6102, Australia}

\author{Philip A. Bland}
\affiliation{School of Earth and Planetary Sciences, Curtin University, Bentley, WA 6102, Australia}

\author[0000-0002-8240-4150]{Martin C. Towner}
\affiliation{School of Earth and Planetary Sciences, Curtin University, Bentley, WA 6102, Australia}

\author[0000-0001-9226-1870]{Hadrien A. R. Devillepoix}
\affiliation{School of Earth and Planetary Sciences, Curtin University, Bentley, WA 6102, Australia}

\author{Martin Cup\'ak}
\affiliation{School of Earth and Planetary Sciences, Curtin University, Bentley, WA 6102, Australia}

\author[0000-0002-5864-105X]{Robert M. Howie}
\affiliation{School of Earth and Planetary Sciences, Curtin University, Bentley, WA 6102, Australia}

\author[0000-0002-0363-0927]{Trent Jansen-Sturgeon}
\affiliation{School of Earth and Planetary Sciences, Curtin University, Bentley, WA 6102, Australia}

\author[0000-0001-9308-0123]{Morgan A. Cox}
\affiliation{School of Earth and Planetary Sciences, Curtin University, Bentley, WA 6102, Australia}

\author[0000-0002-8646-0635]{Benjamin A. D. Hartig}
\affiliation{School of Earth and Planetary Sciences, Curtin University, Bentley, WA 6102, Australia}

\author{Jonathan P. Paxman}
\affiliation{School of Civil and Mechanical Engineering, Curtin University, Bentley, WA 6102, Australia}

\author{Gretchen Benedix}
\affiliation{School of Earth and Planetary Sciences, Curtin University, Bentley, WA 6102, Australia}

\author{Lucy Forman}
\affiliation{School of Earth and Planetary Sciences, Curtin University, Bentley, WA 6102, Australia}

\begin{abstract}
On the 27th of November 2015, at 10:43:45.526 UTC, a fireball was observed across South Australia by ten Desert Fireball Network observatories lasting $6.1\,s$. A $\sim37$ kg meteoroid entered the atmosphere with a speed of 13.68$\pm0.09\,\mbox{km s}^{-1}$ and was observed ablating from a height of 85 km down to 18 km, having slowed to 3.28$\pm0.21 \,\mbox{km s}^{-1}$. Despite the relatively steep 68.5$^\circ$ trajectory, strong atmospheric winds significantly influenced the darkfight phase and the predicted fall line, but the analysis put the fall site in the centre of Kati Thanda - Lake Eyre South. Kati Thanda has metres-deep mud under its salt-encrusted surface. Reconnaissance of the area where the meteorite landed from a low flying aircraft revealed a 60 cm circular feature in the muddy lake, less than 50 m from the predicted fall line. After a short search, which again employed light aircraft, the meteorite was recovered on the 31st December 2015 from a depth of 42 cm. Murrili is the first recovered observed fall by the digital Desert Fireball Network (DFN). In addition to its scientific value, connecting composition to solar system context via orbital data, the recover demonstrates and validates the capabilities of the DFN, with its next generation remote observatories and automated data reduction pipeline.
\end{abstract}

\section{Introduction} \label{sec:intro}
When large meteoroid material, typically of asteroidal origin, encounters the Earth's atmosphere, the bright ($<-4$ mag) phenomenon observed is known as a fireball. 
These are much rarer than fainter meteors which are usually associated with cometary dust. 
Meteoroids that survive the atmosphere are able to deliver asteroidal and even planetary material to Earth as meteorites on the ground. Knowing the origin of a meteorite can give us an understanding not only of composition and formation conditions for a given region of the Solar System, but also assess future impact hazards. 
To determine a likely orbit, the entry of the body must be observed with high spatial and temporal precision from multiple locations. 
Of the thousands of meteorites recovered around the world, less than 40 have orbits associated with them. Dedicated multi-station networks, such as the Desert Fireball Network in Australia, are designed to observe a large area of the sky to capture fireball events with the precision required to calculate orbits. 
The Desert Fireball Network (DFN) was established based on a trial system of 4 film cameras in the Nullarbor Desert \citep{bland2012} that were in operation from 2007 - 2012. After 2 successful recoveries (BR: \citealt{2012M&PS...47..163S}; MG: \citealt{dyl2016characterization}), a digital expansion commenced in 2013 \citep{2017ExA...tmp...19H}. With systems optimised for high resolution and low production cost, 50 autonomous observatories were installed across Western and South Australia over a 2 year period.
Also during this time, an automated data reduction pipeline was developed to increase the efficiency of data collection and processing. This includes the automated detection of fireballs \citep{towner2019fireball}, calibration, triangulation and trajectory analysis of all events. 
The 2.5 million $km^2$ double-station viewing area of the DFN makes it the largest coverage of any fireball network in the world, increasing the likelihood of capturing a meteorite-producing fireball. Its location across the desert regions of the Australian outback also provides favourable conditions and terrain to aid meteorite recoveries. 

On the 27th of November 2015, at 10:43:45.526 UTC, a $6.1$ second fireball was observed across South Australia by ten Desert Fireball Network observatories (Table \ref{table:stations}). The event was triangulated and a mass of $\sim2$ kg was predicted to have landed on Kati Thanda -- Lake Eyre South. This paper details the analysis of the fireball observations, the recovery of the Murrili meteorite and determination of its heliocentric orbit.

\section{Fireball observation and trajectory data}

\subsection{Photographic Data}\label{sec:data}

\begin{figure}
    \centering
    \includegraphics[width=0.8\textwidth]{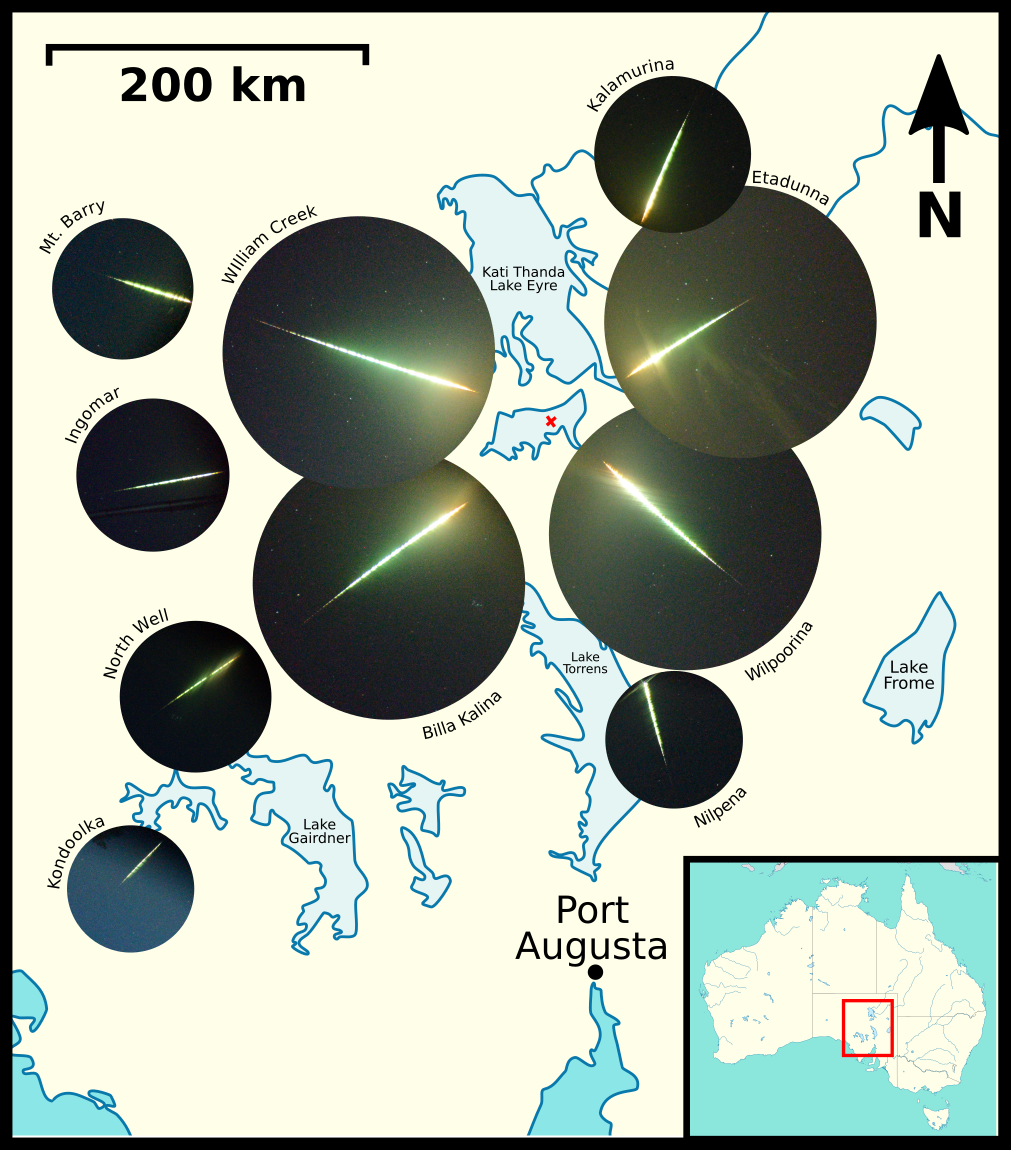}
    \caption{Cropped all-sky images of the fireball from the ten DFN observatories. Images are of the same pixel scale, 
    with the centre of each image positioned at the observatory location on the map. Dashes encoded in the trajectories are an expression of the liquid crystal shutter modulation and provide both absolute and relative timing along each trajectory. Location of the recovered meteorite on Kati Thanda--Lake Eyre South is shown by the red cross.\\\\\\\\\\\\\\}
    \label{fig:images_on_map}
\end{figure}

Ten Desert Fireball Network camera systems imaged the event internally referenced as \textit{DN151127\_01} (Table \ref{table:stations}). The location of these are mapped in Figure \ref{fig:images_on_map} along with the fireball as observed by each system.

The DFN observatories are designed to capture a long exposure image every 30 seconds. When this fireball was recorded, the observatories were configured to capture 25 seconds exposures at ISO 6400 and an aperture setting of F/4. (More recent configuration details can be found in \citet{2017ExA...tmp...19H}.) 
Exposures are triggered roughly simultaneously across the network, however the absolute and relative timing information of the meteoroid flight is encoded in the the long exposure fireball trail itself via a liquid crystal (LC) shutter. This LC shutter is installed between the lens and image sensor in each observatory and is used to modulate the incoming light (by alternating between opaque and translucent states) over time according to a de Bruijn sequence (for details see \citep{2017M&PS...52.1669H, Howie2019timeencoding}). The modulation used at the time of this fireball was pulse width modulation operating at 10 sequence elements per second which produces 20 data points per second accurate to $\pm0.4\,ms$ \citep{2017M&PS...52.1669H}. The operation of the LC shutters is synchronised across the (now global) network via GNSS (Global Navigation Satellite Systems) which allows the DFN to perform individual point wise triangulation.

The identification of the data points produced by the encoded de Bruijn sequence are recorded with the aid of the extraction tool described of \citet{2017ExA...tmp...19H}. This is performed on component (RGB) channels (usually Green) rather than the full colour images seen in Figure \ref{fig:images_on_map}, to reduce the effects of chromatic aberration and saturation in certain channels. 
  
At the time of fireball observation, the majority of cameras experienced near perfect sky conditions, with some clouds low on the horizon for Mt. Barry, Etadunna and Nilpena stations only. 
The William Creek camera was the first to image the fireball at 10:43:45.526 UTC on the 27th of November 2015, with the Wilpoorinna system capturing the final observation at 10:43:51.626 UTC. 
Despite saturation on closer cameras, particularly Etadunna due to added cloud brightness,
122 points were identified along the trajectory across all systems -- corresponding to the entire de Bruijn sequence over this 6.1 second observable flight (see additional material for reduced data). The observation angles were highly favourable, especially of the four closest cameras which were near equidistant to the event. 

\begin{table}[!h]
	\caption{Locations of Desert Fireball Network Observatories that obtained photographic records of the event. Times are relative to first fireball observation at 10:43:45.526 UTC on the 27th of November 2015 (from the William Creek camera). 
	We separate stations that are $<200\,$km from the event as this denotes a limit of astrometric precision.}              
	\label{table:stations}      
	\center                                      
	\resizebox{\textwidth}{!}{
	\begin{tabular}{l c c c| c c c}         
		\hline\hline                       
		 & & Observatory && range\tablenotemark{*} &  start\tablenotemark{$\dagger$} time & end\tablenotemark{$\dagger$} time \\
		name & latitude & longitude & altitude (m)  & (km) & observed & observed \\
		\hline      
		William Creek  &  28.91566 S & 136.33495 E & 79  & 121 & 0.00 & 5.86 \\
		Wilpoorinna  &  29.96502 S & 138.31090 E & 91  & 141 & 0.06 & 6.10 \\ 
		Billa Kalina &  30.23769 S & 136.51565 E & 114  & 145  & 0.06 & 5.92 \\
		Etadunna & 28.72019 S & 138.65290 E & 29 & 162  & 0.20 & 5.86 \\
		\hline
		Kalamurina & 27.75920 S & 138.23471 E & 0 & 204 & 0.90 & 5.7   \\
		Nilpena & 31.02330 S & 138.23257 E &112 & 224 & 0.20 & 4.06 \\
		Ingomar & 29.58553 S & 135.03868 E & 197 & 250 & 1.1 & 5.7 \\
        Mount Barry & 28.51652 S & 134.88627 E & 173 & 259 & 0.90 &  5.46  \\
		North Well & 30.85765 S & 135.27432 E & 176 & 270 & 1.1  & 4.96   \\
		Kondoolka & 31.98066 S & 134.84909 E & 252 & 388 & 1.3 &  4.7 \\ 
		\hline                                            
	\end{tabular}}
		\tablenotetext{*}{distance from the observatory to the meteoroid at 65 km altitude}
		
\end{table}

\newpage
\subsection{Astrometry}\label{sec:astrometry}

From pixel coordinates, the positions of the meteoroid in each image are astrometrically calibrated using the background star field. This corrects for distortions of the fish eye lens used and mitigates issues of atmospheric refraction.
Despite the addition of the LC shutter and settings optimised for bright fireballs, DFN images are capable of imaging stars down to 7.5  point source limiting magnitude. Images taken within close temporal proximity to a fireball event are used to calibrate the lens for a given system, and map pixel coordinates to astrometric observations in azimuth and elevation.

\begin{figure}[h!]
    \centering
    \includegraphics[width=0.5\textwidth]{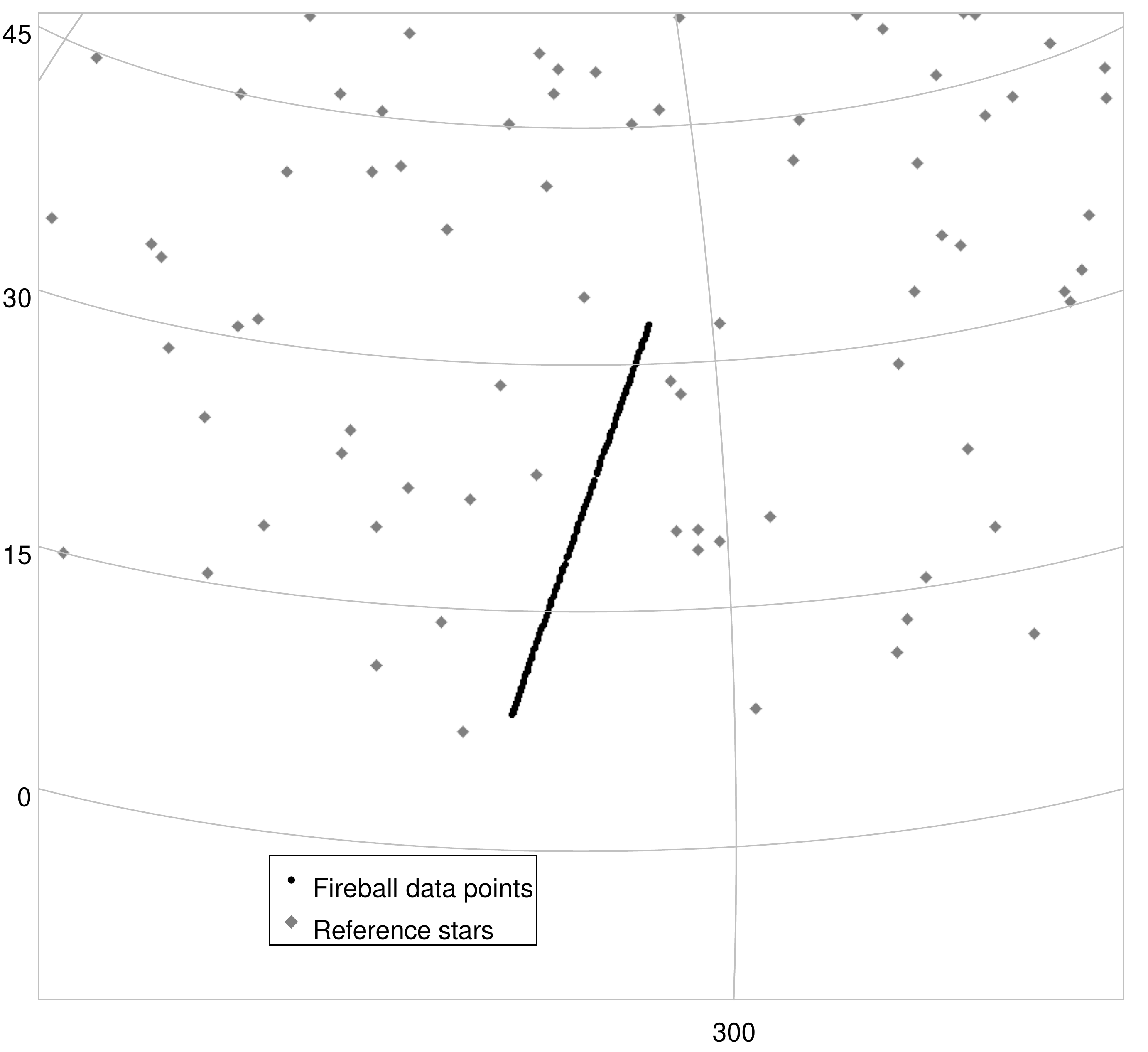}
    \caption{Horizontal coordinates plot of the Wilpoorinna fireball data points (observation epoch 2015-11-27T10:43Z), and a subset of the calibration stars (observation epoch 2015-11-27T10:50Z). This shows that the lowest fireball data point ( azimuth $\lambda = 313.626\degr \pm 0.028$ East of North and geometric altitude $\phi = 8.573\degr \pm0.031 $) is still within the convex hull of reference stars, which allows precise error control over the whole trajectory. \\\\
    }
   
    \label{fig:astrom_fireball_ref_stars}
\end{figure}

\begin{figure}[h!]
    \centering
    \includegraphics[width=0.5\textwidth]{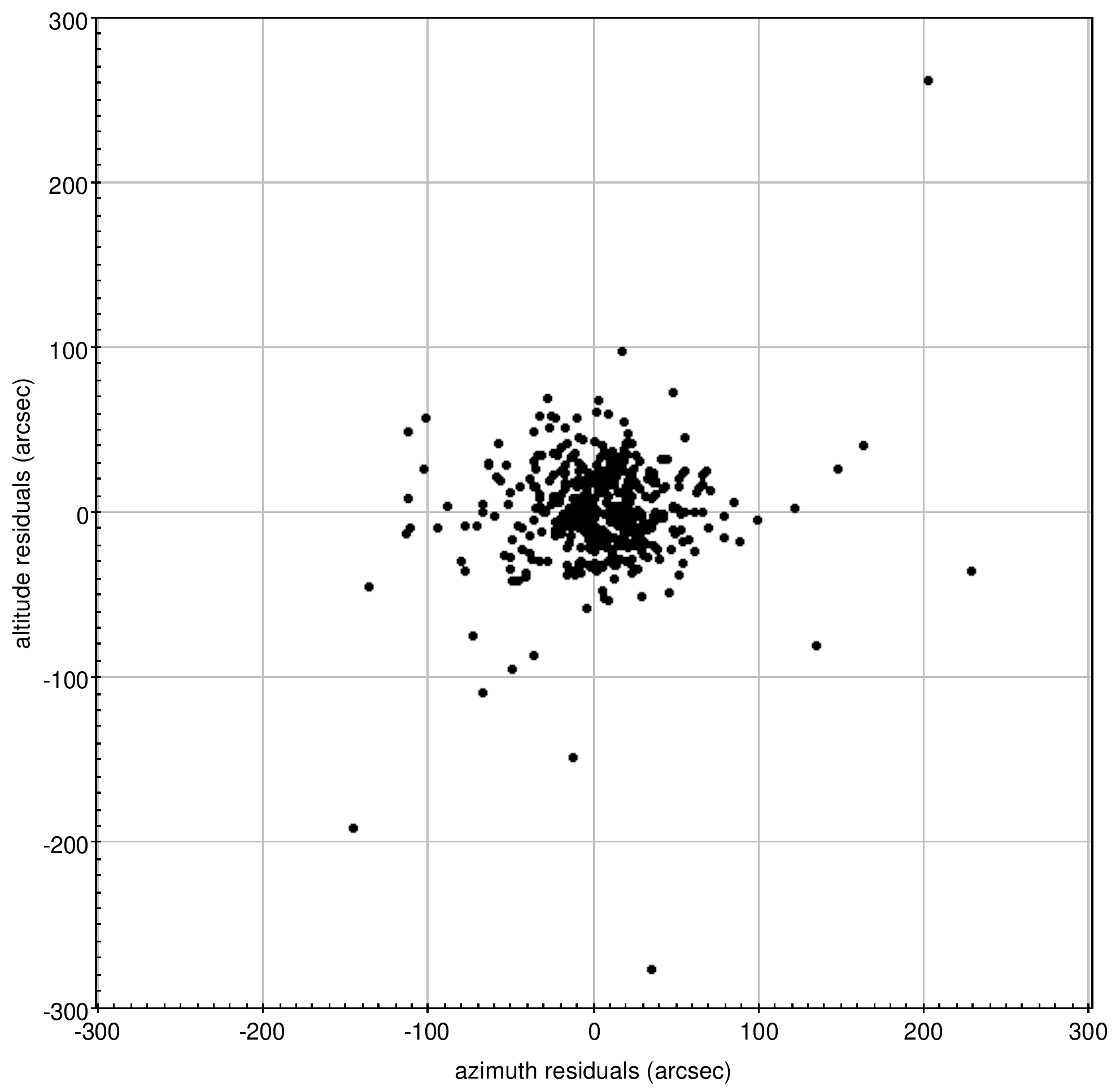}
    \caption{Residuals on the global astrometric fit for the Wilpoorinna calibration frames (412 stars).\\\\}
    \label{fig:astrom_global_fit_residuals}
\end{figure}

Here we detail the calibration procedure, using as example the Wilpoorinna camera data.
The meteoroid entered during astronomical twilight -- shortly after sunset, but just before a 96\% illuminated moon rise. For any given event, a calibration image is automatically chosen to acquire a fame taken during astronomical night, with no moon or cloud cover. In this case, the software chose an image taken a few days later to satisfy these conditions. This is admissible as camera systems are incredibly stable - though only during a more recent test by \citet{devillepoix2018dingle} was this quantified, showing less than a $0.017^{\circ}$ variation in pointing angle over a 3 month period.
As this was still untested at the time of the event, to ensure maximum astrometric accuracy we manually selected calibration exposures. For Wilpoorinna, this was just 6 minutes after the fireball, giving similar atmospheric conditions.
This came at the expense of a brighter sky background due to twilight, resulting in only 412 reference stars (instead of the some 1000 on an image taken during astronomical nighttime) used to compute the global astrometric solution from this camera (Fig. \ref{fig:astrom_fireball_ref_stars} and \ref{fig:astrom_global_fit_residuals}).
Despite this, we are still able to get precise astrometry down to 5\degr above the horizon thanks to the clear, non-polluted skies the Australian outback offers.
Although Murrili fell in a relatively poorly covered area of the network (all cameras observed the event from over $100$ km), we were able to get to nominal arcminute astrometric precision for all data points on the 4 closest cameras ($<200$ km range; Figure \ref{table:stations}; raw astrometric data available in supplementary material, with raw image NEFs available at DOI:10.5281/zenodo.3891468).

\newpage
\section{Trajectory Modelling}\label{sec:modelling_and_orbit}

\subsection{Initial Triangulation} \label{sec:triang}
\begin{figure}[!h]
	\centering
	\includegraphics[width=\textwidth]{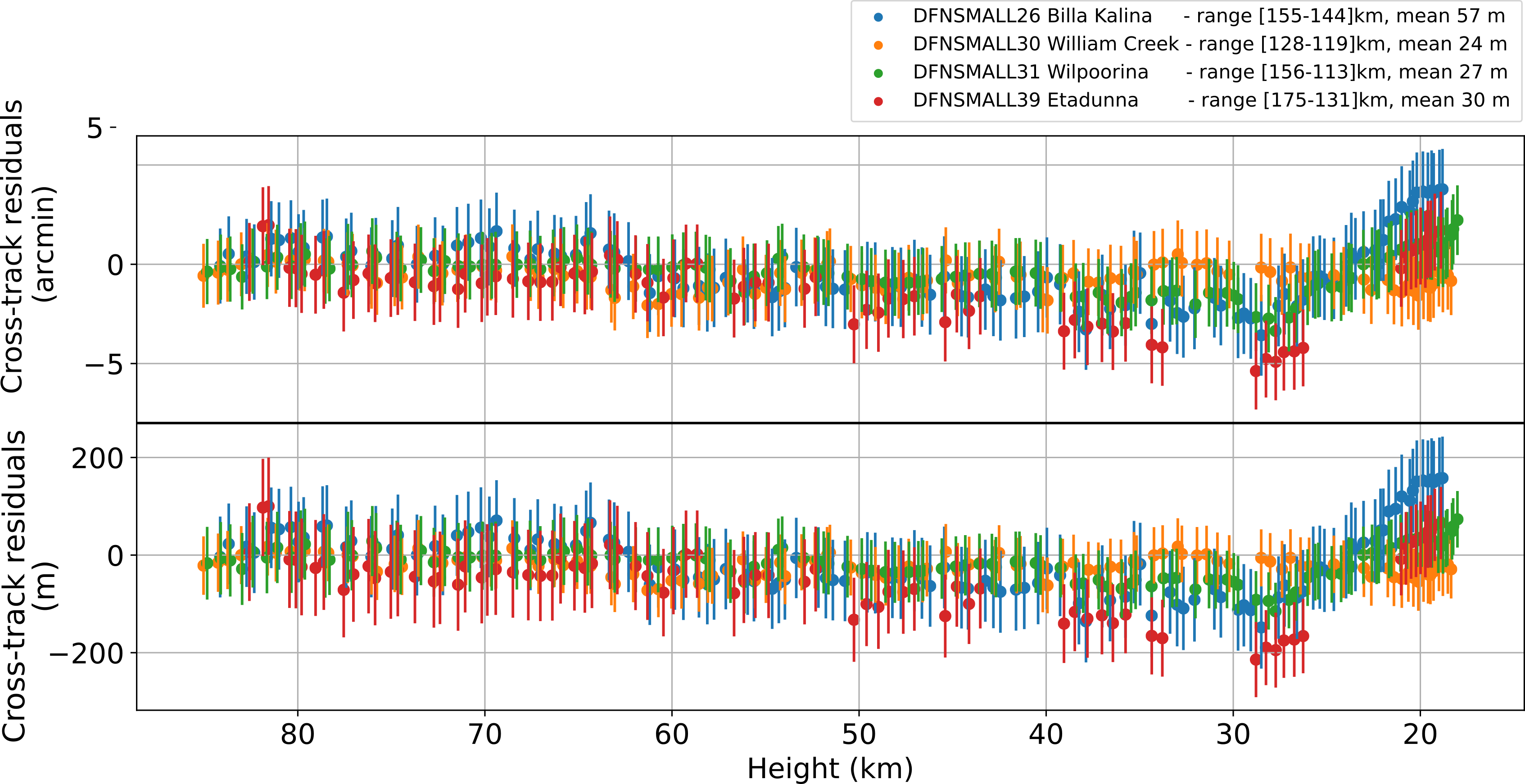}
	\caption{Cross-track residuals of the straight line least squares fit to the trajectory from each view point (top: in angular residuals. bottom: in distance projected on a perpendicular plane to the line of sight). The error bars displayed correspond to the 1-$\sigma$ uncertainties on the astrometry.
	Ranges given in the legend are from the camera to the [highest - lowest] trajectory point. Mean values are given for the absolute residuals for each camera, with an overall mean of 30 m for all cameras. Note that orbit calculations use a }
	\label{fig:cross_track_residuals}
\end{figure}

An approximate trajectory of the observed fireball can be initially triangulated using the straight line least squares (SLLS) method of \citet{1990BAICz..41..391B} in an inertial reference frame. This assumes a straight line trajectory and creates a radiant in 3D space that minimises the angular residuals to the observed lines of sight, with observatory locations corrected for Earth rotation effects.
The convergence angle of observation planes between the four closest observatories is near ideal: 79\degr/81\degr/68\degr/49\degr (East of North from Etadunna.
Although all camera observations are used in this triangulation, the angular uncertainties are magnified by distance from the event, 
resulting in high cross-track residuals for cameras with lower astrometric precision and of increasing range (Fig. \ref{fig:cross_track_residuals}). The resulting apparent radiant from this full trajectory fit has a right ascension and declination of $\alpha=337.35^\circ$,  $\delta=-29.43^\circ$ respectively (J2000).

Although cross-track errors provide an assessment of SLLS fit, there is no quantification of the three dimensional error in the trajectory without performing further analyses. This triangulation method is also unable to incorporate errors associated with using a straight line assumption. Without a full assessment of both observational and model errors, uncertainties will be underestimated. This will significantly affect fall estimates and orbital parameter calculations, exacerbating errors within their respective integration techniques.

\begin{figure}[!h]
	\centering
	\includegraphics[width=0.8\textwidth]{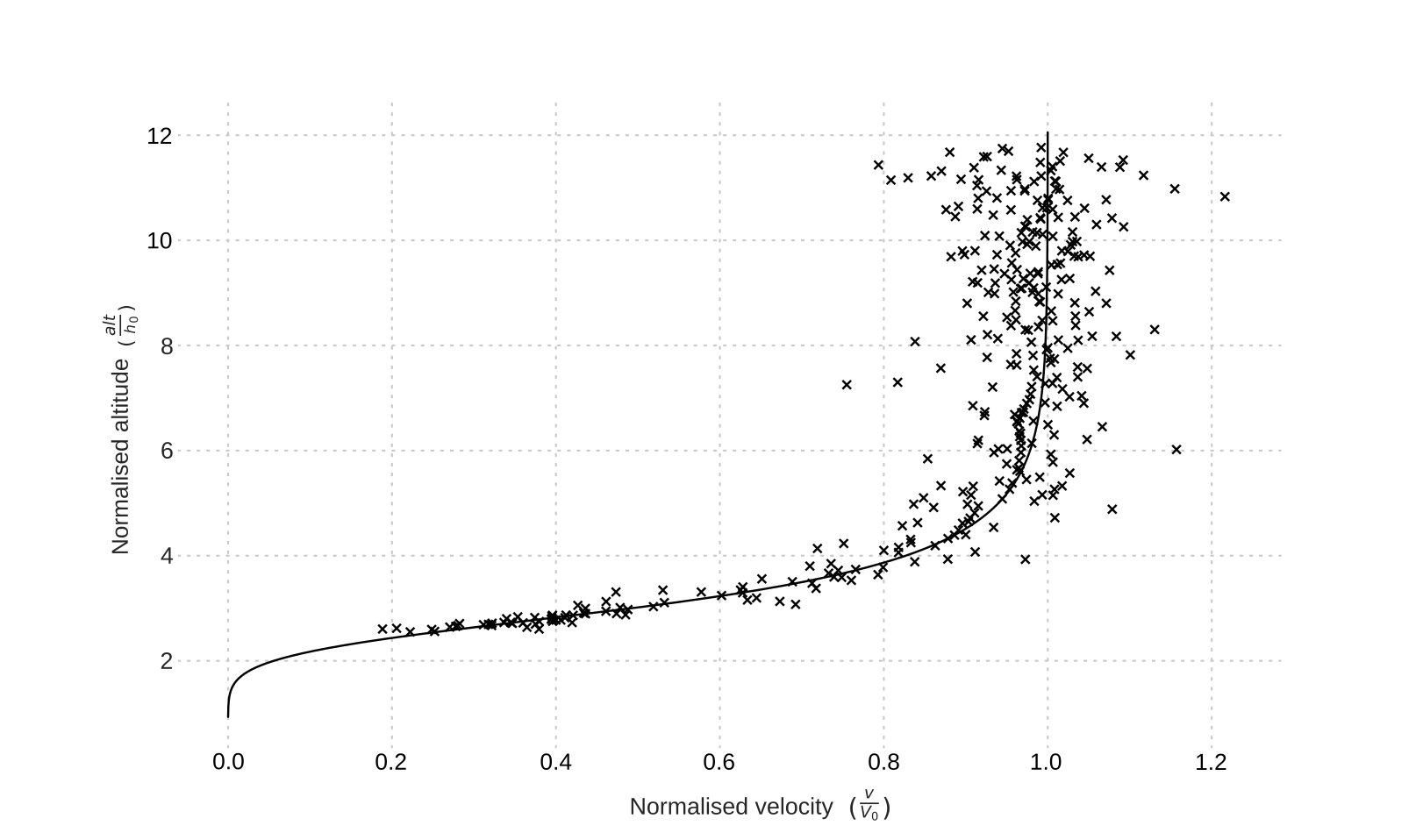}
	\caption{Trajectory data with velocities normalised to the velocity at the top of the atmosphere ($V_0$=13.68 $\mbox{km s}^{-1}$; Tab. \ref{tab:traj_sum}) and altitudes normalised to the atmospheric scale height, $h_0 = 7.16$ km. The best fit to Equation 10 of \citet{2009AdSpR..44..323G} results in $\alpha=8.34$ and $\beta=1.06$ and is shown by the black line.}
	\label{fig:alphabeta}
\end{figure}

From this straight line approximation, the fireball trajectory was observed to begin at an altitude of $84$ km, with a $68^\circ$ angle to the local horizontal at a velocity of $14.18\mbox{km s}^{-1}$.
The final observation was made at a height of $18$ km with a meteoroid velocity of $3.37\,\mbox{km s}^{-1}$, having flown a $73.01$ km long trajectory.
We use the $\alpha$--$\beta$ criterion to determine if the fireball is a likely meteorite-dropping candidate \citep{gritsevich2012,sansom2019determining}. The dimensionless ballistic ($\alpha$) and mass loss ($\beta$) parameters calculated for this fireball are $\alpha=8.34$ and $\beta=1.063$ (Figure \ref{fig:alphabeta}). It is viable to use this as a first pass, and allowed us to establish that there was likely a terminal mass to recover. These values predict a significant fall and further modelling is required to determine the likely meteorite mass. 

\subsection{Estimating initial and terminal masses}\label{sec:eks_grits}
With this first order approximation of the trajectory, we use an extended Kalman filter and smoother to estimate the initial and terminal masses of the meteoroid. 
This method (outlined by \citet{2015M&PS...50.1423S}) estimates the meteoroid \textit{state} -- position and velocity along a straight line, as well as the main mass -- according to single body aerodynamic theory \citep{Hoppe1937, 1971JGR....76.4653B, Stulov1995}.
A \textit{filter} runs unidirectional
in time, considering only previous data to improve the current state estimate. As fireball data are not acquired in real time, a \textit{smoother} is subsequently run in reverse to allow all data to influence every timestep. 
The filter is initiated at $t_f=6.1s$, with initial state values for position$=73.01\pm1.5$ km, velocity$=3.4\pm0.5\mbox{km s}^{-1}$ and mass$=1\pm1$ kg. 
As the terminal mass of a meteoroid will be low compared to its initial mass, it is a more highly constrained filter input and the reason we initiate the filter at the terminal observation point. 
For this first-order, simplified model, an assumed final mass close to nil is used as the filter input (and is subsequently updated by the 'reverse' smoother phase). Other assumed parameters include values for meteoroid characteristics including shape (set to be a sphere; $A=1.21$), density ($\rho_m=3500\,\mbox{kg m}^{-3}$), aerodynamic drag coefficient\footnote{$\Gamma$ is referred to as the drag factor in many meteoroid trajectory works, including \citep{Ceplecha2005} and is related to the aerodynamic drag such that $c_d = 2\Gamma$ \citep{Bronshten1983, 2015aste.book..257B}.} 
($c_d=1$) and ablation coefficient\footnote{As we are including the process of fragmentation, this is the more commonly termed apparent ablation coefficient.} ($\sigma=0.014\,\mbox{s}^{2}\,\mbox{km}^{-2}$; \citealt{Ceplecha2005}).
The filter predicts changes to the state (position, velocity and mass) using the single body aerodynamic equations \citep{2015M&PS...50.1423S}, and then updates the values with an observation -- here the distance from $t_0$ along the straight line trajectory. Although the uncertainties in the model are large to account for assumptions and unmodelled phenomena such as fragmentation, they are incorporated -- independent to observation errors. 
This allows a more comprehensive investigation of the final uncertainties in state values. 

The initial mass estimated using this method gives $m_0=37.9\pm2.3$ kg and final mass $m_f=1.9\pm0.4$ kg, with velocities $v_0=13.68\pm0.09\,\mbox{km s}^{-1}$ and $v_f=3.28\pm0.21\,\mbox{km s}^{-1}$, and total trajectory length $72.11\pm0.04$ km.
These mass values are consistent with the initial and terminal masses calculated using the method of \citet{Gritsevich2007morp, 2009AdSpR..44..323G} for the same characteristic assumptions and the $\alpha$--$\beta$ parameters given in Section \ref{sec:triang} where $m_0=31$ kg and, assuming rapid enough spin for consistent ablation on all surfaces, $m_f=1.5$ kg. 

\begin{table}[!h]
	\centering
	\begin{tabular}{lll}
		\hline
		& Beginning & Terminal \\
		\hline
		Time (UTC) & 10:43:45.526 & 10:43:51.626\\
		Longitude (WGS84; $\degr$E) &137.20841&137.47817\\
		Latitude (WGS84; $\degr$S) &29.29583 &29.26534\\
		Height (km) &  $84.97  \pm  0.02$ & $17.96 \pm0.04$ \\
		Velocity ($\mbox{km s}^{-1}$) & $13.68\pm0.09$ &  $3.28\pm0.21$ \\
		Angle from local horizontal ($^\circ$)& 68.5 & 68.2 \\
		Apparent entry radiant (RA $^\circ$)& 337.38$\pm$0.01& --\\
		Apparent entry radiant (Dec $^\circ$) & -29.38$\pm$0.01& --\\
		Calculated mass\footnote{Assumptions listed in Section \ref{sec:eks_grits}} (kg) & 38$\pm$2 &  $1.9\pm0.4$   \\
		\hline
		Total duration & 6.1 s \\
		Total length & $72.11\pm0.04$ km\\
		Recovery location (WGS84)&137.537650$^\circ$ E & 29.260890$^\circ$ S \\
		Recovered mass & 1.68 kg &\\
		Bulk density of meteorite \footnote{\citet{macke2016density}}& $3470\pm10\,\mbox{kg m}^{-3}$&\\
		\hline
	\end{tabular}
	\caption{Summary table of fireball parameters using extended Kalman filter (EKS) modelling on a straight line assumption fitted to the whole trajectory, along with recovered meteorite properties. Note that apparent radiant values are calculated using the upper trajectory only, as described in Section \ref{sec:orb}.
	}
	\label{tab:traj_sum}
\end{table}
\pagebreak[1]
\newpage
\section{Radiant and Heliocentric orbit determination}\label{sec:orb}
As stated in Section \ref{sec:modelling_and_orbit}, the straight line trajectory may not be relied upon to provide representative entry vectors of the meteoroid from which to calculate orbits. Both the radiant and velocity magnitude when using the whole dataset may be skewed by aerodynamic effects of the decelerating body \citep{ellie20173dpf}, while the uncertainties without considering both model and observational errors in 3D will be underestimated.  Figure \ref{fig:cross_track_residuals} highlights a skew in the residuals toward the end of the trajectory, indicating a deviation from the straight line assumption, possibly from unmodelled aerodynamic effects.
An entry velocity calculated using the EKS will provide a more rigorous analysis of the errors on its magnitude, though will be large with high observational errors associated with using pre-triangulated SLLS positions. To reduce these, observations prior to significant deceleration are isolated and a straight line assumption may be more valid. Recalculating the initial velocity with only observations above 60 km in this case gives identical results ($v_0=13.68\pm 0.09)$, though apparent radiant (J2000) does change: $\alpha=337.38\pm0.01^\circ$,  $\delta=-29.38\pm0.01^\circ$ (values given as best result in summary Table \ref{tab:traj_sum}). This is a 0.056$^\circ$ or 3.4 arcminute difference to the full trajectory fit in Section \ref{sec:data}. This is beyond the one sigma error, showing the influence of underlying model assumptions (such as a straight line trajectory) that are often overlooked in fireball trajectory analyses. 

\begin{figure}[h!]
    \centering
    \includegraphics{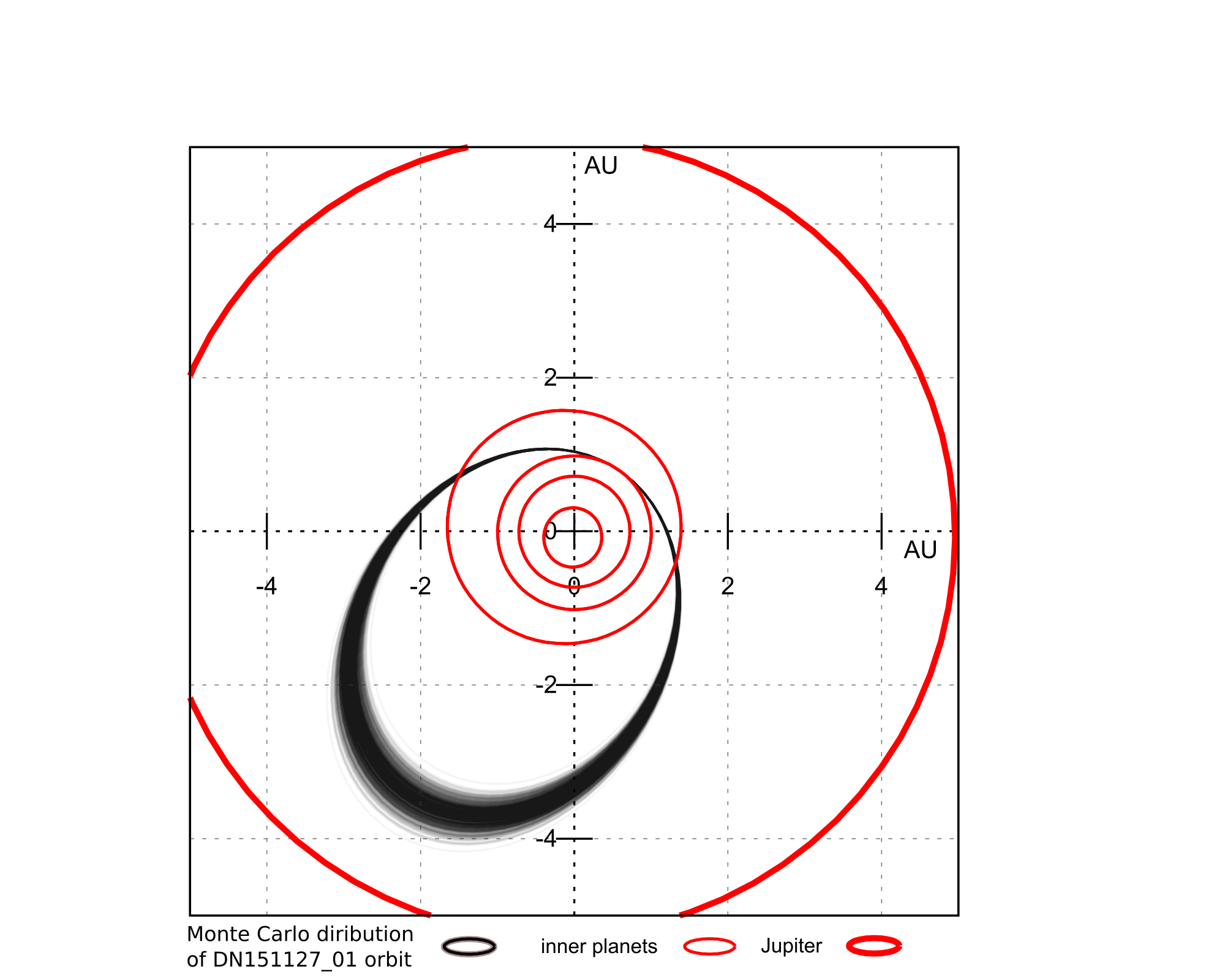}
    \caption{Monte Carlo orbit regression analysis for the \textit{DN151127\_01} meteoroid}
    \label{fig:orb}
\end{figure}

These apparent radiant values are used to calculate the 
pre-atmospheric orbit of the meteoroid using the integration method of \citet{2019M&PS...54.2149J}. The Monte Carlo results are illustrated in Figure \ref{fig:orb}, with orbital values given in Tab. \ref{tab:orbit}. This low inclination orbit, close to the 3:1 mean-motion resonance with Jupiter is quite common for an H5 chondrite \citep{2014me13.conf...57J}.

\begin{table}[!h]
    \centering
    \begin{tabular}{ccc}
    \hline 
    Epoch & TDB & 2015-11-27\\
  Semi-major axis $a$ & AU & 2.521 $\pm$ 0.075 \\
  Eccentricity $e$ & &  0.609 $\pm$ 0.012 \\
  Inclination $i$ & \degr   & 3.32 $\pm$ 0.060 \\
   Argument of perhelion $\omega$ & \degr  & 354.557 $\pm$ 0.039 \\
    Longitude of the Ascending Node $\Omega$ & \degr  &   64.7420 $\pm$ 0.0033 \\
    Perihelion distance $q$ & AU & 0.9851196  $\pm$ 0.000006  \\
     Aphelion distance  $Q$ & AU & 4.06 $\pm$ 0.15 \\
     Corrected radiant (RA) $\alpha _g$ & \degr   & 330.68 $\pm$ 0.15 \\
     Corrected radiant (Dec) $\delta _g$ & \degr  & -28.561 $\pm$ 0.027 \\
     Geocentric speed $V _g$ & $\mbox{m s}^{-1}$ &  8245 $\pm$ 142 \\
     Tisserant parameter wrt. Jupiter $T _J$ & & 3.16 \\
    \hline 
    \end{tabular}
    \caption{Pre-encounter orbital parameters expressed in the heliocentric ecliptic frame (\textit{J2000}) and associated $1\sigma$ formal uncertainties.}
    \label{tab:orbit}
\end{table}

\newpage
\section{Darkflight and wind modelling}

\begin{figure}[!h]
    \centering
    \includegraphics[width=0.7\textwidth]{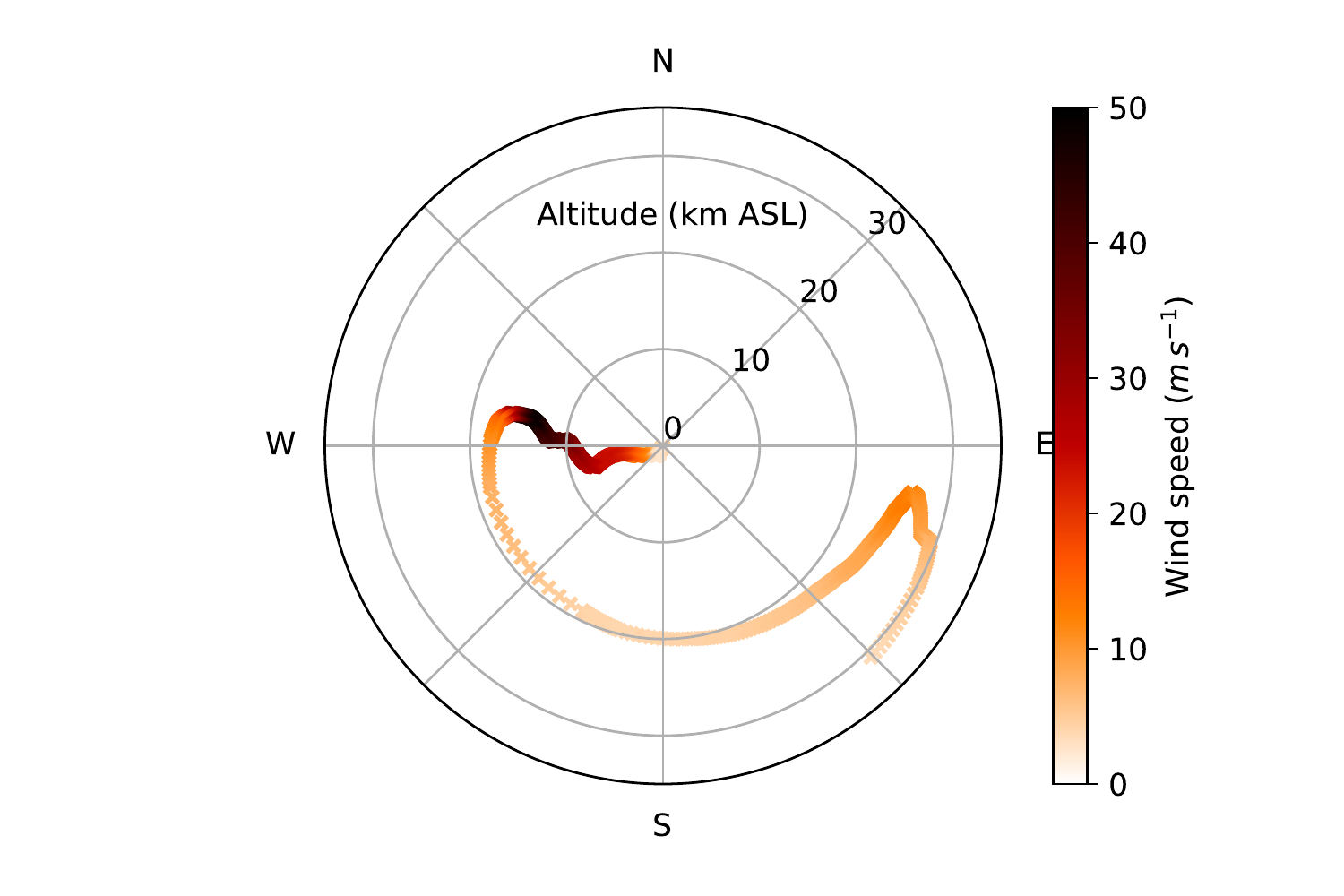}
    \caption{Wind model (speed and direction for a given altitude), extracted as a vertical profile at the coordinates of the lowest visible bright flight measurement for 2015-11-27T10:45Z.}
    \label{fig:Wind}
\end{figure}

The final observation point of the fireball was at an altitude of $17.96 \pm0.04$ km. The dark flight trajectory of the meteorite is substantially affected by the atmospheric winds. In central South Australia, where the Murrili meteorite fell, the subtropical jet stream is typically predominant. The atmospheric winds were numerically modelled using the Weather Research and Forecasting (WRF) model version 3.7.1 with dynamic solver ARW (Advanced Research WRF) \citep{skamarock2008description}. The weather modelling was initialised using global data with 1 degree resolution from the National Centers for Environmental Prediction (NCEP) Final analysis (FNL) Operational Model Global Tropospheric Analysis data. From this, a 1-km resolution WRF was produced with a 15 minute history interval, and a weather profile for 2015-11-27T10:45Z was then extracted at the final position of the luminous flight. The weather profile (Figure \ref{fig:Wind}) includes wind speed, wind direction, pressure, temperature and relative humidity at heights ranging up to 30 km, fully covering the dark flight altitudes.

\begin{figure}
    \centering
    \includegraphics[width=0.6\textwidth]{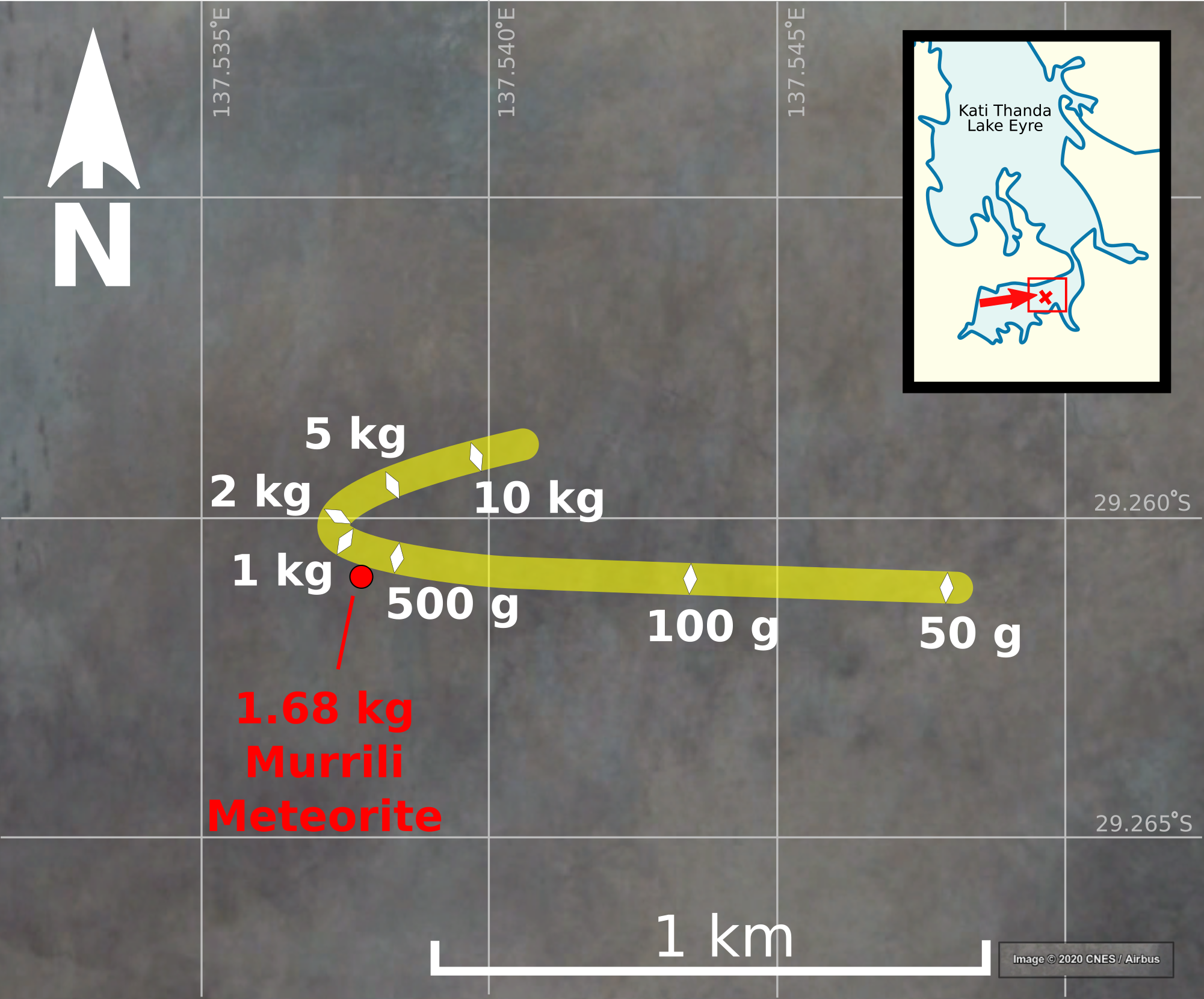}
    \caption{Fall area lies within the southern region of Kati Thanda (insert). Fall line in yellow shows positions for spherical masses ($A=1.21$). The location of the recovered meteorite (red dot; $\lambda = 137.5376\degr$ $\phi = -29.2609\degr$ (WGS84).) is $<50$ m from the fall line, and $\simeq 100\,\mbox{m}$ from its corresponding spherical mass position. }
    \label{fig:fall_line}
\end{figure}

The strongest wind affected altitudes around $13.5\,$km and exceeded $45\,\mbox{m s}^{-1}$ with a westerly direction (270-280\degr). This significantly influenced the fall of the meteorite, shifting the low mass end of the fall line East (Figure \ref{fig:fall_line}).

\section{Search and recovery}

At the time of the event, the DFN operated 18 cameras in South Australia, 5 of them without network connection.
Of the 4 closest cameras (Table \ref{table:stations}), only Billa Kalina and Wilpoorinna had internet connection, this was enough to determine that a large meteorite on the ground was likely, and also which stations needed to be visited to get extra close viewpoints (William Creek and Etadunna).

\begin{figure}[h!]
	\centering
	\includegraphics[width=0.4\textwidth]{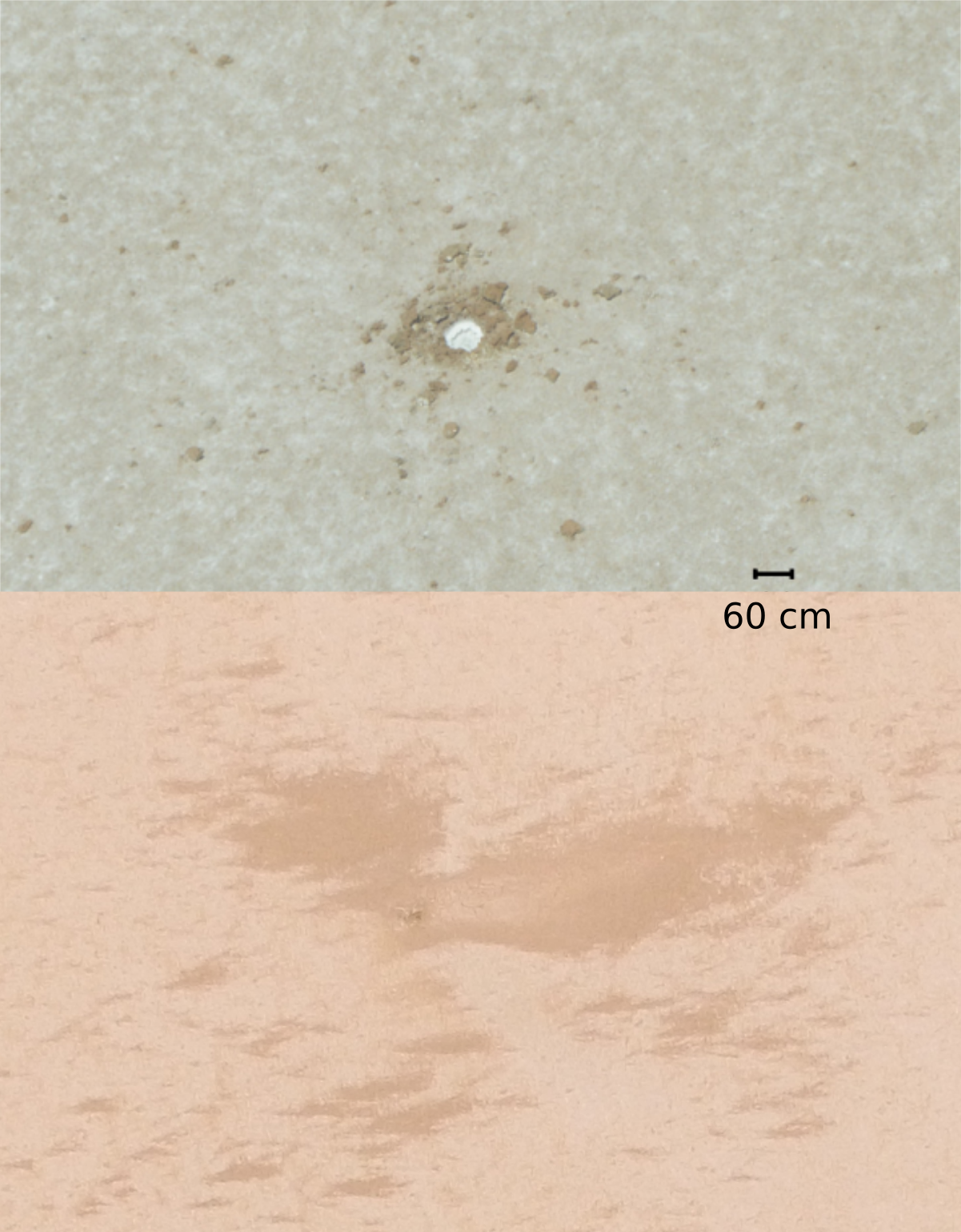}
	\caption{Images taken of the `splash zone' from (top) light aircraft on December 15th 2015 and (bottom) drone on December 30 2015. Note rainfall between these dates infilled the hole and significantly reduced the distinctive features of the impact.}
	\label{fig:splash}
\end{figure}

\begin{figure}
    \centering
    \includegraphics[width=0.8\textwidth]{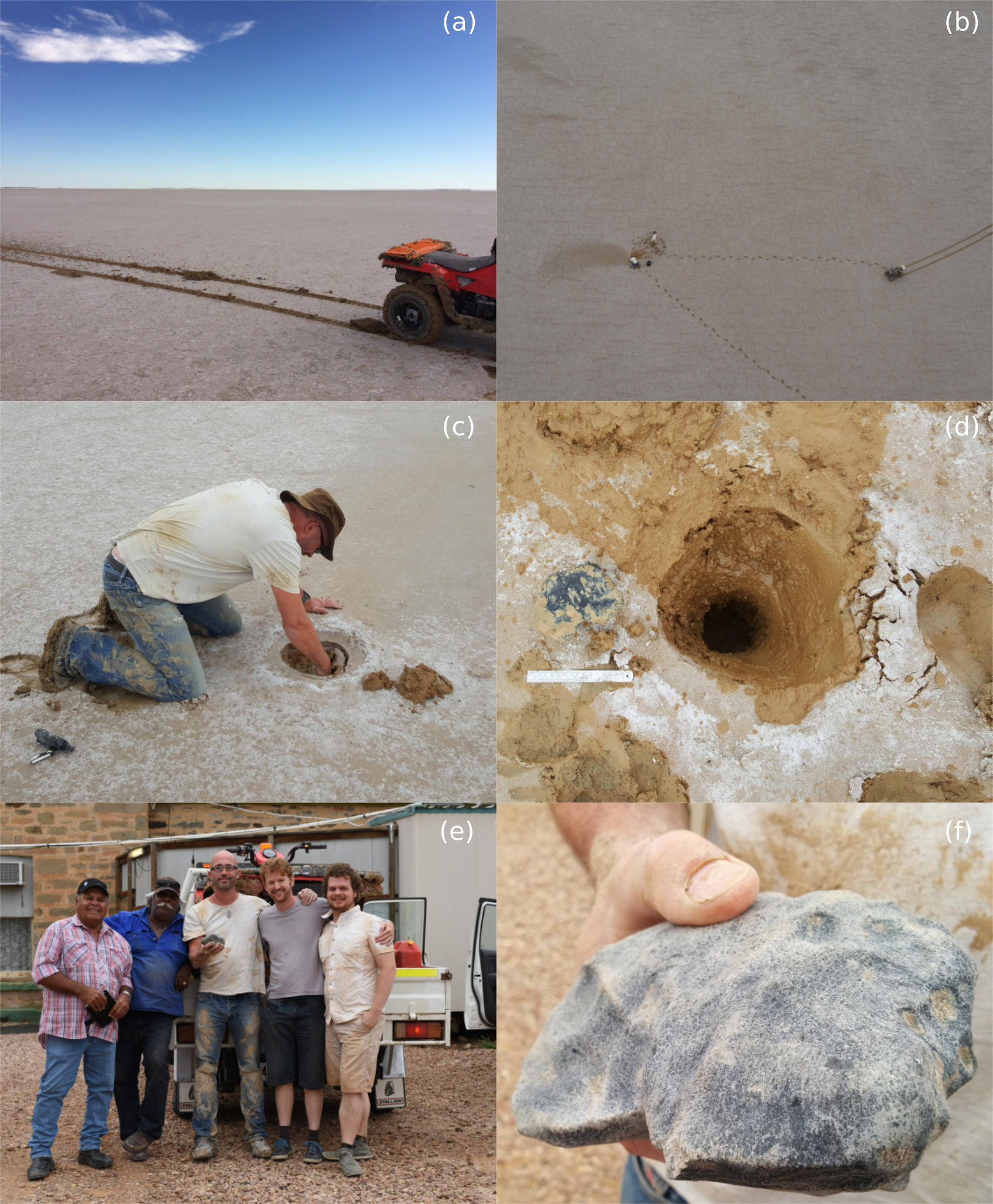}
    \caption{Images of the Murrili meteorite recovery. (a) The challenging salt lake surface; (b) target acquired--image from light aircraft of the approach and PB and RH digging at the impact site; (c) PB excavating the soft mud infill; (d) Murrili caked in mud after being removed from a 42 cm deep hole, ruler is 15 cm; (e) joyous search team outside the Marree Hotel; (f) Murrili meteorite shortly after recovery and removal of mud, fusion crust is fully intact. Images available at \url{https://commons.wikimedia.org/wiki/File:PB_holding_Murrili_Meteorite.jpg} under a Creative Commons Attribution-ShareAlike 4.0 International.}
    \label{fig:collage}
\end{figure}

While on this expedition, the team (MC, BH) managed to get onto an unplanned scenic flight over the lake to take a look, just in case something was visible from the sky.
This revealed a suspicious 'splash' (Figure \ref{fig:splash} top) in the lake close to the initial predicted impact point, although it was not possible to acquire precise coordinates with the equipment used.
This feature was later estimated to be $\simeq60\,cm$ in diameter, using the approximate altitude of the aircraft and the focal length of the camera used. As it was the only feature of this nature visible, it is unlikely that any other significant mass landed and is the main remaining mass of the meteoroid. 
After reduction of all available data (see section \ref{sec:triang}), some time was taken in arranging the appropriate permissions required to travel on Kati Thanda.
The area is a place of significance for the Arabana people, the indigenous traditional owners of Kati Thanda and the surrounds. 
A search was coordinated from the town of Marree, located approximately 70 km southeast of the predicted fall site where a number of residents were even witness to the fireball event.
The search party consisted of three members of the research team from Curtin University (PB, RH, JP), and two guides from the Arabana people, Dean Stuart and Dave Strangways.
The team was equipped with three four wheel drive vehicles and two single passenger all-terrain vehicles (quad bikes). 
26.4 mm of rain on the 22nd of December (Bureau of Meterorology at Marree) gave rise to uncertain stability on the lake surface.
An initial reconnaissance trip on December 29th determined that the lake was dry on the surface, but that movement was challenging due to a thick layer of mud just under the salt crust (Figure \ref{fig:collage}a). 
This lead to a quad bike and foot search by two members, around the best estimate from the plane reconnaissance and fall line predictions. As the clearly defined splash had been washed away, and with the meteorite being buried, it was near impossible to locate from the ground. A small Unmanned Aerial Vehicle (UAV) (Figure \ref{fig:splash} bottom) and a further manned aerial search were able to re-locate the site and direct the ground team (Figure \ref{fig:collage}b). This was very close to the original predicted fall line (Figure \ref{fig:fall_line}). All that remained of the impact impression in the mud was a small depression at coordinates $\lambda = 137.5376\degr$ $\phi = -29.2609\degr$ (WGS84) (Figure \ref{fig:collage}c).
The mud at this spot was distinctly softer, and after digging 42 cm into the mud, revealed the Murrili meteorite (Figure \ref{fig:collage}). The particular area of Kati Thanda on which this meteorite was recovered is named after Elder Murrili, giving the meteorite its name\footnote{application submitted to Meteoritical Bulletin Database and approved 31 Mar 2016}. The fusion crust was complete with some striations across the surface (Figure \ref{fig:collage}f). This heart-shaped rock was measured to be 13 x 7 x 6 cm, weighing $1.68$ kg with a bulk density of $3470\pm10\,\mbox{kg m}^{-3}$ \citep{macke2016density}, and later classified by Gretchen Benedix at Curtin University to be an H5 ordinary chondrite \citep{benedix2016mineralogy, metbul_murrili}. The distinctly non-spherical shape of the Murrili meteorite easily accounts for the shift along the predicted fall line from corresponding spherical mass assumptions \mbox{(Figure \ref{fig:fall_line})}.

The fast recovery of this rock after a 3 day search campaign was paramount, as 16.6 mm of rain fell on the 1st of January 2016, along with significant rain to the lake feeder regions, led to Kati Thanda filling up over the next few days, and to a 30 year record high in the following months.
This prevented any further searching for fragments.

\section{Conclusions}
As the first meteorite recovered using the digital expansion of the Desert Fireball Network, the successfully recovery of the Murrili meteorite tested and validated all aspects of the newly developed hardware and software pipelines. The fireball on the 27th of November 2015, at 10:43:45.526 UTC was observed by 10 DFN observatories. Four of these were within 200 km of the trajectory and allowed a high quality observational dataset to be recorded. A terminal mass was predicted to have fallen close to a `splash' observed in lake Kati Thanda -- Australia's largest salt lake. After rain this feature had been erased, yet a small team successfully recovered the 1.68 kg rock at a depth of 42 cm. 
Simple modelling of the fireball trajectory predicted a terminal mass consistent with the meteorite recovered. From an early reconnaissance flight, only one impact impression was seen. It is therefore reasonable to assume that this was the only significant mass to recover,parameter $\sim 1$ 
and modelling of a single body mass was suitable for this event. Weather restrictions prevented the further search for fragments.
The calculated orbit of this H5 ordinary chondrite is aligned with the 3:1 mean motion resonance with Jupiter. A full petrological analysis of the Murrili meteorite will be presented in a forthcoming publication.

\section*{Acknowledgements}
The authors would like to thank the Arabana people for their help and involvement in recovering the Murrili meteorite, in particular Dean Stuart and Dave Strangways, as well as Trevor Wright of Wrightsair for piloting and providing the aircraft for the aerial reconnaissance flights which detected the fall location their support in conducting aerial searches.
The authors would also like to thank the Marree Hotel for their hospitality as well as the hosts of our DFN cameras, in particular the William Creek township, Wilpoorina Station, Billa Kalina Station, Etadunna Station, the Australian Wildlife Conservancy for use of the Kalamurina Wildlife Sanctuary, Nilpena Station, Ingomar Station, Mount Barry Station, North Well Station and Koondoolka Station.
This work was funded by the Australian Research Council as part of the Australian Discovery Project scheme (DP170102529), and receives institutional support from Curtin University. Data reduction is supported by resources provided by the Pawsey Supercomputing Centre with funding from the Australian Government and the Government of Western Australia.
The DFN data reduction pipeline makes intensive use of Astropy, a community-developed core Python package for Astronomy \citep{2013A&A...558A..33A}.
Supplemental data are provided, with raw images made available by \citet{images}.
\bibliography{biblio}
\bibliographystyle{abbrvnat}

\end{document}